  \documentclass[10pt]{article}

  \usepackage[english]{babel}
  \usepackage[utf8]{inputenc}

  \usepackage[T1]{fontenc}          %

   \usepackage{geometry}
  \usepackage{url}
  \usepackage{graphicx}
  \usepackage{float}
  \usepackage{xcolor}
  \usepackage{amsmath}
  \usepackage{amssymb}

  \numberwithin{equation}{section}
  \numberwithin{figure}{section}

  \usepackage[unicode=true, backref=false, colorlinks=true]{hyperref}
  \hypersetup{linkcolor=blue, citecolor=orange, urlcolor=magenta}

\begin{document}

\rightline{\today}
{~}
\vspace{.5cm}

\begin{center}

{\LARGE \bf (Non-)uniqueness of Einstein-Palatini Gravity  }
\vspace{.5cm}

{\large \bf
  Bert Janssen,${}^{a,b}$
  Alejandro Jim\'enez-Cano,${}^{a,b}$ \\
  Jos\'e Alberto Orejuela,${}^{a,b}$ 
  Pablo S\'anchez-Moreno${}^{c,d}$ 
}\footnote{E-mail addresses: bjanssen@ugr.es, alejandrojc@ugr.es, josealberto@ugr.es, pablos@ugr.es}

\vspace{.2cm}
{\it 
  ${}^a$Departamento de F\'isica Te\'orica y del Cosmos \\
  ${}^b$Centro Andaluz de F\'isica de Part\'iculas Elementales \\
  ${}^c$Departamento de Matem\'atica Aplicada \\
  ${}^d$Instituto Carlos I de F\'isica Te\'orica y Computacional
}
\vspace{.1cm}

{\it
  Facultad de Ciencias, Avda Fuentenueva s/n,\\
  Universidad de Granada, 18071 Granada, Spain
}

\vspace{1truecm}
\end{center}

\begin{center} {\bf ABSTRACT} \end{center}

\vspace{2mm}

\noindent
We analyse the most general connection allowed by Einstein-Hilbert theory in Palatini formalism. We also consider a matter lagrangian independent of the affine connection. We show that any solution of the equation of the connection is essentially Levi-Civita up to a term that contains an undetermined 1-form. Finally, it is proved that these connections and Levi-Civita describe a completely equivalent physics. \\

\noindent {\footnotesize This is a preprint of the following work: B. G. Sidharth, J. Carnicer, M. Michelini, C. Perea (Editors), {\it Fundamental Physics and Physics Education Research}, 2020, Springer, reproduced with permission of Springer Nature Switzerland AG. \\ The final authenticated version is available online at: \url{http://dx.doi.org/10.1007/978-3-030-52923-9} }

\quad

\noindent
{\footnotesize Talk given by A.J.C. at the workshop Frontiers of Fundamental Physics 15 (Orihuela, Spain, November 2017).}

\section{Introduction and mathematical notions}

Since the publication of the Einstein's theory of General Relativity in 1915, we understand gravitation as a geometrical effect. Many extensions of this theory have been formulated in order to solve various problems in theoretical physics, such as dark matter or the first corrections to General Relativity that could come from the quantum gravity regime.

In the geometrical framework introduced by Einstein, the \emph{spacetime} is defined as a differentiable manifold $\mathcal{M}$. Omitting some mathematical details, a $D$-\emph{dimensional manifold} is essentially a topological space that looks, locally, as the Euclidean space $\mathbb{R}^{D}$. For example, spheres, planes and hyperboloids are 2-dimensional manifolds.

Additionally, we include a \emph{lorentzian metric tensor}, $g_{\mu\nu}$, which allows to measure lengths, volumes and so on. Hence it is possible to talk about the module of a vector that is not necessarily non-negative, due to the lorentzian signature. Those vectors that are not trivial but have zero norm determine the lightlike paths and, then, light cones that define the casual structure of the spacetime. 

\begin{figure}[H]
\centering{}
  \includegraphics[width=0.2\paperwidth]{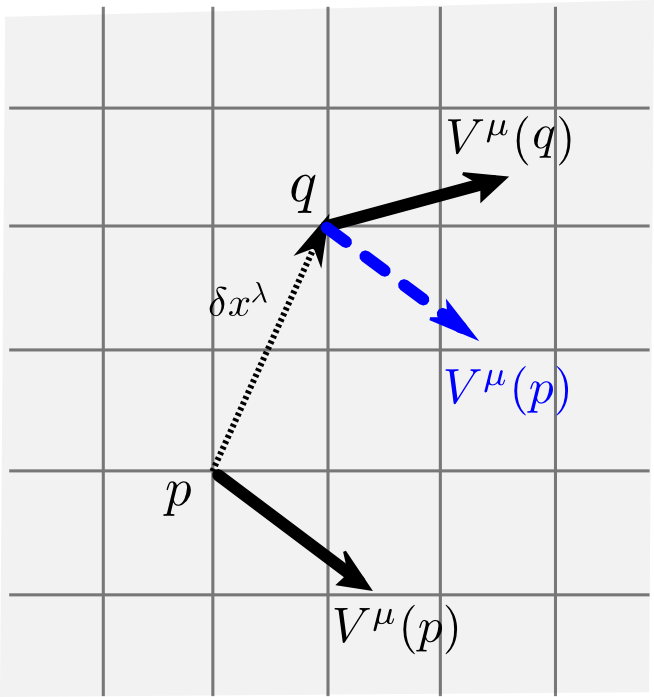}
  \caption{\label{fig1}\textit{Comparison of vectors in Euclidean space (natural notion of parallelism).}}
\end{figure}

Another fundamental notion that can be defined, even in the absence of metric, is \emph{parallelism}. The motivation for this additional concept is the following. Consider the Euclidean space $\mathbb{R}^{D}$ and a couple of vectors in different points, $p$ and $q$ [Figure \ref{fig1}]. If we want to compare them, we simply take, for example, the one in $p$ and move it to $q$ keeping the vector parallel to itself and without changing the module. And, finally, we subtract both vectors to see the difference.

\begin{figure}[H]
\centering{}
  \includegraphics[width=0.4\paperwidth]{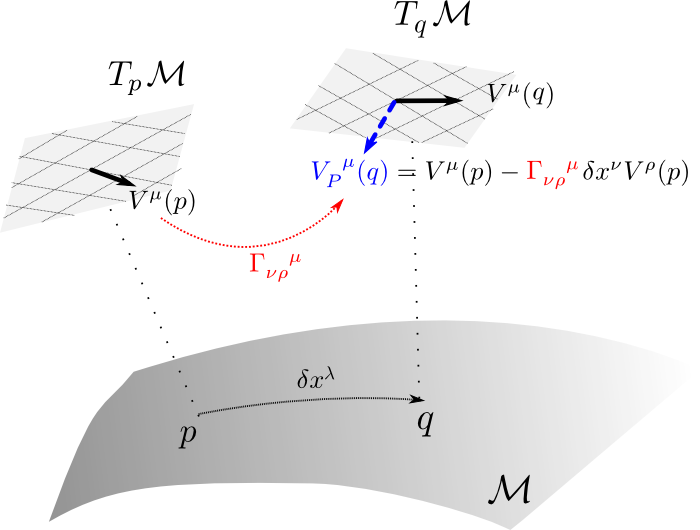}
  \caption{\label{fig2}\textit{An affine connection represents a notion of parallel in the manifold and allow to parallely transport vectors along curves. The dashed line is, by definition, the parallel transport of $V^{\mu}(p)$ from $p$ to $q$ along the $x^{\lambda}$ direction. This vector and $V^{\mu}(q)$ can be compared since both live in $T_{q}\mathcal{M}$.}}
\end{figure}

However, if the manifold is general, the initial vectors live in different spaces (the tangent spaces at $p$ and $q$, respectively, $T_{p}\mathcal{M}$ and $T_{q}\mathcal{M}$) and there is no natural way to relate them [Figure \ref{fig2}]. In the Euclidean space both tangent spaces can be identified making the comparison trivial. In the general case, we need to introduce an additional structure that carries the information about parallelism, the \emph{affine structure}, whose fundamental object is the \emph{(affine) connection} \emph{$\Gamma_{\mu\nu}{}^{\sigma}$}. Once we have a connection, given a curve between two points, we have a rule to relate vectors in them: the \emph{parallel transport} associated to the connection.

The connection permits the definition of a \emph{covariant derivative} ('covariant' means that once applied to a tensor, the result is also a tensor), and other intrinsic geometrical properties of the spacetime, such as the \emph{curvature} and the \emph{torsion}, respectively:
\begin{align}
R_{\mu\nu\rho}{}^{\lambda} & \equiv\partial_{\mu}\Gamma_{\nu\rho}{}^{\lambda}-\partial_{\nu}\Gamma_{\mu\rho}{}^{\lambda}+\Gamma_{\mu\sigma}{}^{\lambda}\Gamma_{\nu\rho}{}^{\sigma}-\Gamma_{\nu\sigma}{}^{\lambda}\Gamma_{\mu\rho}{}^{\sigma}\,,\\
T_{\mu\nu}{}^{\lambda} & \equiv\Gamma_{\mu\nu}{}^{\lambda}-\Gamma_{\nu\mu}{}^{\lambda}\,.
\end{align}

Consider a manifold with a metric structure $g_{\mu\nu}$, then it can be proved that there is only one connection, $\Gamma^{(g)}{}_{\mu\nu}{}^{\lambda}$, compatible with the metric and torsionless, namely
\begin{equation}
\nabla_{\lambda}g_{\mu\nu}=0\,,\qquad T_{\mu\nu}{}^{\lambda}=0\,.
\end{equation}
This connection is called the \emph{Levi-Civita connection} of $g_{\mu\nu}$, and it is completely determined by the metric:
\begin{equation}
\Gamma^{(g)}{}_{\mu\nu}{}^{\lambda}=\frac{1}{2}g^{\lambda\sigma}\left(\partial_{\mu}g_{\sigma\nu}+\partial_{\nu}g_{\mu\sigma}-\partial_{\sigma}g_{\mu\nu}\right)\,.
\end{equation}
In fact, given a metric, the Levi-Civita affine structure is the simplest connection we can deal with. The metric compatibility and the nullity of the torsion simplify many geometrical identities. Moreover, we are not introducing extra degrees of freedom in the theory, just the ones that come from the metric.

Before starting with the physics, let us introduce a few useful definitions. A \emph{curve} $\gamma(\alpha)$ is a differentiable function $\gamma\,:\,I\rightarrow\mathcal{M}$, where $I$ is an interval of the real line. The image of a curve in the manifold is what we will call \emph{trajectory} or \emph{path}. So a trajectory is a set of spacetime points joined in a continuous and differentiable way, while the curve is the function that generates this set. 

Let $\mathcal{M}$ be an spacetime equipped with a connection $\Gamma_{\mu\nu}{}^{\rho}$. An \emph{autoparallel} of this affine structure is the image of a curve whose velocity is parallel to itself (with respect to $\Gamma_{\mu\nu}{}^{\rho}$). In other words, given a curve $\gamma(\alpha)$ with velocity $v^{\mu}(\alpha)$, its image is an autoparallel if the following equation holds:
\begin{equation}
v^{\mu}\nabla_{\mu}v^{\rho}\equiv\frac{\mathrm{d}v^{\rho}}{\mathrm{d}\alpha}+\Gamma_{\mu\nu}{}^{\rho}v^{\mu}v^{\nu}=f(\alpha)v^{\rho}
\end{equation}
for some function $f(\alpha)$. If we reparametrize the trajectory, by doing $\alpha\rightarrow\beta(\alpha)$, we change the velocity as well as the function $f$. This function can always be set to zero (identically) with an appropriate choice of the parameter. Those are called \emph{affine parameters} for the trajectory.

Consider the manifold also has a metric structure. The autoparallels of the associated Levi-Civita connection are special because they can be derived from a completely metric approach, i.e. without using the Levi-Civita parallelism. If the velocity is timelike or spacelike, they correspond to trajectories that are critical points of the length functional:
\begin{equation}
s[\gamma](\alpha)=\int_{0}^{\alpha}\sqrt{\left|g_{\mu\nu}v^{\mu}v^{\nu}\right|}\mathrm{d}\alpha'\,.
\end{equation}

The lightlike case should be treated separately, because the length functional cannot be varied smoothly. However they can be seen as critical points of other functionals that, again, only involve the metric structure. For these reasons, in general, we will call the autoparallels of Levi-Civita \emph{critical trajectories}. The associated affine parameters have a special meaning, because their changes are proportional to the length between the considered points. In the timelike case, it is essentially the \emph{proper time}, so these parameters represent the rhythm of a proper (i.e. freely falling) clock along them.

\section{Einstein's equations and variational principles}

General relativity is a geometric theory of the spacetime whose dynamics is described by the \emph{Einstein's equation}:
\begin{equation}
R_{\mu\nu}(g)-\frac{1}{2}g_{\mu\nu}R(g)=-\kappa\mathcal{T}_{\mu\nu}\,,\label{EinsEq}
\end{equation}
where $R_{\mu\nu}(g)\equiv R_{\mu\lambda\nu}{}^{\lambda}(g)$ is the Levi-Civita Ricci tensor, $R(g)\equiv g^{\mu\nu}R_{\mu\nu}(g)$ is the Levi-Civita Ricci scalar, $\kappa\equiv8\pi G$ ($G$ is the Newton' s constant) and $\mathcal{T}_{\mu\nu}$ is the (Hilbert) \emph{energy-momentum tensor} that contains the information about the matter and energy content.

This equation can be obtained from a more fundamental object through a variational principle, the \emph{Einstein-Hilbert action}: 
\begin{equation}
S[g,\,\psi]=\frac{1}{2\kappa}\int R(g)\,\sqrt{\left|g\right|}\mathrm{d}^{D}x\,\,+\,\,S_{\text{matter}}[g,\,\psi]\,.
\end{equation}
The energy-momentum tensor is then defined by
\begin{equation}
\mathcal{T}_{\mu\nu}\equiv\frac{2}{\sqrt{\left|g\right|}}\frac{\delta S_{\text{matter}}}{\delta g^{\mu\nu}}\,.
\end{equation}

Notice that we are assuming (from the start) a particular affine structure, the one fixed by the metric. We are selecting the Levi-Civita connection by hand and this can be considered natural because it is the simplest one. When we obtain from a gravitational action the equations of motion admitting that the affine structure is Levi-Civita, we are using the so-called \emph{metric formalism}, because the metric determines everything related to the gravitational field.

Another approach, which is called \emph{Palatini} or \emph{metric-affine formalism}, consists in considering the metric and the connection as independent fields. Now the connection is general and the corresponding equations of motion should determine whether it is Levi-Civita or not. The action in this formalism is
\begin{equation}
S[g,\,\Gamma,\,\psi]=\frac{1}{2\kappa}\int g^{\mu\nu}R_{\mu\nu}(\Gamma)\,\sqrt{\left|g\right|}\mathrm{d}^{D}x\,\,+\,\,S_{\text{matter}}[g,\,\psi]\,.
\end{equation}
It is worth remarking that we are assuming that the matter part of the action does not depend on the affine connection $\Gamma$.

This formalism is interesting because we expect Levi-Civita connection to be fixed by the corresponding equations of motion, in contrast with the metric formalism in which it is selected artificially. 

\section{Palatini solutions of the Einstein-Hilbert action}

If we vary the Einstein-Hilbert action in the metric-affine formalism we obtain the following equations of motion for the metric and the
connection, respectively:
\begin{align}
0 & =\frac{1}{2}\Bigl(R_{\mu\nu}(\Gamma)+R_{\nu\mu}(\Gamma)\Bigr)-\frac{1}{2}g_{\mu\nu}R(\Gamma)+\kappa\mathcal{T}_{\mu\nu}\,,\label{metricEq}\\
0 & =\nabla_{\lambda}g^{\mu\nu}-\nabla_{\sigma}g^{\sigma\nu}\delta_{\lambda}^{\mu}+\frac{1}{2}g^{\mu\nu}g^{\rho\tau}\nabla_{\lambda}g_{\rho\tau}-\frac{1}{2}g^{\rho\tau}\nabla^{\nu}g_{\rho\tau}\delta_{\lambda}^{\mu}\nonumber \\
 & \qquad+T_{\lambda\sigma}{}^{\sigma}g^{\mu\nu}+T_{\rho\sigma}{}^{\sigma}g^{\rho\nu}\delta_{\lambda}^{\mu}+T_{\sigma\lambda}{}^{\mu}g^{\sigma\nu}\,.\label{connectEq}
\end{align}

The equation \eqref{connectEq} can be simplified if the dimension of the spacetime is $D>2$.\footnote{For the particular case $D=2$ see \cite{Deser}.} We then obtain:

\begin{equation}
0=\nabla_{\lambda}g_{\mu\nu}-T_{\nu\lambda}{}^{\sigma}g_{\mu\sigma}+\frac{1}{D-1}T_{\lambda\sigma}{}^{\sigma}g_{\nu\mu}+\frac{1}{D-1}T_{\nu\sigma}{}^{\sigma}g_{\lambda\mu}\,.\label{connectEqSimp}
\end{equation}
Clearly, Levi-Civita is a solution, because in that case each term of the right hand side vanishes. However let us try for other solutions. Consider only those that are torsionless, then, necessarily, $\nabla_{\lambda}g_{\mu\nu}$ should be zero, so Levi-Civita is the only possibility. The same happens for metric-compatible solutions. In fact, when Palatini formalism is presented (in textbooks for example) one of these two conditions is assumed. Consequently, we lose the information about the general solution and it reduces to Levi-Civita.

The most general solution of the equation \eqref{connectEqSimp} has the form:
\begin{equation}
\Gamma_{\mu\nu}{}^{\sigma}=\Gamma^{(g)}{}_{\mu\nu}{}^{\sigma}+\mathcal{A}_{\mu}\delta_{\nu}^{\sigma}\,,
\end{equation}
where $\mathcal{A}_{\mu}$ is an arbitrary 1-form \cite{Bernal}. We will call it Palatini connection from now on. The associated covariant derivative of the metric (also called \emph{non-metricity} tensor) and torsion are:
\begin{equation}
\nabla_{\lambda}g_{\mu\nu}=-2\mathcal{A}_{\lambda}g_{\mu\nu}\,,\qquad T_{\mu\nu}{}^{\sigma}=\mathcal{A}_{\mu}\delta_{\nu}^{\sigma}-\mathcal{A}_{\nu}\delta_{\mu}^{\sigma}\,.
\end{equation}
Here, we clearly notice what we stated before: switching off one of them implies $\mathcal{A}_{\lambda}=0$ and, then, Levi-Civita as the only possibility.

The Palatini Riemann tensor, Ricci tensor and Ricci scalar are given by
\begin{align}
R_{\mu\nu\rho}{}^{\lambda}(\Gamma) & =R_{\mu\nu\rho}{}^{\lambda}(g)+\mathcal{F}_{\mu\nu}\delta_{\rho}^{\lambda}\,,\\
R_{\mu\nu}(\Gamma) & =R_{\mu\nu}(g)+\mathcal{F}_{\mu\nu}\,,\\
R(\Gamma) & =R(g)\,,
\end{align}
respectively, where $\mathcal{F}_{\mu\nu}=\partial_{\mu}\mathcal{A}_{\nu}-\partial_{\nu}\mathcal{A}_{\mu}$. As a consequence of these expressions, the equation of motion of the metric \eqref{metricEq} becomes the Einstein's equation \eqref{EinsEq}, see \cite{DadhichPons}. We now present some properties of these solutions.

\subsubsection*{Projective relation between solutions}

Any two Palatini connections, for example
\begin{equation}
\Gamma_{\mu\nu}{}^{\sigma}=\Gamma^{(g)}{}_{\mu\nu}{}^{\sigma}+\mathcal{A}_{\mu}\delta_{\nu}^{\sigma}\,,\qquad\Gamma'{}_{\mu\nu}{}^{\sigma}=\Gamma^{(g)}{}_{\mu\nu}{}^{\sigma}+\mathcal{A}'{}_{\mu}\delta_{\nu}^{\sigma}\,,
\end{equation}
are related by a transformation:
\begin{equation}
\Gamma_{\mu\nu}{}^{\sigma}\rightarrow\Gamma'{}_{\mu\nu}{}^{\sigma}=\Gamma_{\mu\nu}{}^{\sigma}+k_{\mu}\delta_{\nu}^{\sigma}\,,
\end{equation}
for certain covector $k_{\mu}$. This transformation is a \emph{projective transformation} which means that preserves autoparallels. This can be proved easily. First, consider an autoparallel trajectory for the connection $\Gamma'{}_{\mu\nu}{}^{\sigma}$,
\begin{equation}
\frac{\mathrm{d}v^{\rho}}{\mathrm{d}\beta}+\Gamma'{}_{\mu\nu}{}^{\rho}v^{\mu}v^{\nu}=0\,.
\end{equation}
Then, imposing the projective relation between both connections and defining $-k_{\mu}v^{\mu}\equiv f(\beta)$ we get to the expression:
\begin{equation}
\frac{\mathrm{d}v^{\rho}}{\mathrm{d}\beta}+\Gamma_{\mu\nu}{}^{\rho}v^{\mu}v^{\nu}=f(\beta)v^{\rho}\,.
\end{equation}
And this is the equation of an autoparallel for the connection $\Gamma_{\mu\nu}{}^{\rho}$ with a non-affine parametrization. If we parametrize the path affinely ($\beta\rightarrow\alpha$ and $v^{\rho}\rightarrow u^{\rho}$) we obtain:
\begin{equation}
\frac{\mathrm{d}u^{\rho}}{\mathrm{d}\alpha}+\Gamma_{\mu\nu}{}^{\rho}u^{\mu}u^{\nu}=0\,.\,\,\text{(QED)}
\end{equation}

Consequently,\emph{ the whole set of Palatini connections shares the same autoparallels}, up to reparametrizations, which have no physical meaning. As a matter of fact, since Levi-Civita is a particular Palatini connection (the case with $\mathcal{A}_{\mu}=0$) we conclude: \emph{the autoparallels of any Palatini connection are critical trajectories of the metric}.

\subsubsection*{Homothety property }

Let $\gamma(\tau)$ be a general curve in the spacetime and $v^{\mu}$ its velocity, and a vector $W^{\mu}$. If we compare the change of
the vector along $\gamma(\tau)$ under Palatini and Levi-Civita parallel transport, we see that the difference between both transports is proportional to $W^{\mu}$:
\begin{equation}
\left(v^{\mu}\nabla_{\mu}-v^{\mu}\nabla^{(g)}{}_{\mu}\right)W^{\rho}=-\mathcal{A}_{\mu}v^{\mu}W^{\rho}\equiv\lambda(\tau)W^{\rho}\,.
\end{equation}
The module is not conserved but the direction does. Due to this, we say the Palatini parallel transport is \emph{homothetic with respect to the Levi-Civita transport}. It can be proved that the only connections with this property are the Palatini connections \cite{Bernal}. Any other connection would generate a perturbation in the direction
of $W^{\mu}$.

\section{Observability and physical implications}

We introduced the Palatini formalism in order to see if the dynamics could fix Levi-Civita as the fundamental connection of the theory, in contrast with the metric formalism in which it is selected by hand. However, we have obtained a family of connections that differ in a vector field, with Levi-Civita as a particular case. In this section we analyse the physical implications of this field. Indeed, we will see that it is undetectable or, equivalently, that metric and Palatini formalism describe the same physics.

The main point is that the gravitational dynamics is the same in both formalisms. The equation of the matter is clearly the same, because the corresponding action does not depend on the connection, so the difference between formalisms does not affect this part of the total action. And, as we previously showed, Palatini connections imply the reduction of the equation of the metric to the Einstein's
equation. The resulting dynamics for the metric and the matter content is given by
\begin{equation}
R_{\mu\nu}(g)-\frac{1}{2}g_{\mu\nu}R(g)=-\kappa\mathcal{T}_{\mu\nu}\,,\qquad\frac{\delta S_{\text{matter}}}{\delta\psi}=0\,,
\end{equation}
in both formalisms.

Furthermore, in Einstein-Hilbert gravity the distinction between critical and autoparallel trajectories disappears due to the projective symmetry. In fact, defining the trajectory of a test particle is often presented as a basic problem of metric-affine theories. Those of critical length and those with covariantly constant velocity are candidates because both of them infinitesimally reduce to straight lines. The critical paths are the simplest approach, but there are authors who defend the description with autoparallels \cite{Kleinert} and others who state that only the conserved currents determine the
test paths \cite{Hehl}. 

In Einstein-Hilbert gravity, the conservation of the energy-momentum tensor selects the critical paths. However, fortunately, the autoparallels set by the Palatini dynamics coincide, as a consequence of the projective relation between Palatini connections (a family that includes Levi-Civita). Indeed, we have seen that the field $\mathcal{A}_{\mu}$ can be eliminated by the freely falling observer with an appropriate choice of the parameter. 

All of these ideas point in the same direction: the field $\mathcal{A}_{\mu}$ has no physical effects \cite{Bernal}.

\section{Equivalence in other theories}

Finally, we add a few remarks about Palatini connections in other theories. One example is \emph{Lovelock theory} in Palatini formalism:
\begin{equation}
S[g,\,\Gamma,\,\psi]=S_{\text{Lov}}[g,\,\Gamma]+S_{\text{matter}}[g,\,\psi]\,,\qquad S_{\text{Lov}}[g,\,\Gamma]\equiv\int\sum_{n=1}^{K}a_{n}\mathcal{L}_{n}(g,\,\Gamma)\,\sqrt{\left|g\right|}\mathrm{d}^{D}x\,,
\end{equation}
where $a_{n}$ are certain dimensionful parameters, $\mathbb{N}\ni K\leq\text{ceiling}\left(D/2-1\right)$ and the $n^{\text{th}}$-order Lovelock lagrangian is defined by
\begin{equation}
\mathcal{L}_{n}(g,\,\Gamma)=\delta_{\nu_{1}}^{[\mu_{1}}...\delta_{\nu_{2n}}^{\mu_{2n}]}g^{\rho_{1}\nu_{1}}...g^{\rho_{n}\nu_{2n-1}}R_{\mu_{1}\mu_{2}\rho_{1}}{}^{\nu_{2}}(\Gamma)...R_{\mu_{2n-1}\mu_{2n}\rho_{n}}{}^{\nu_{2n}}(\Gamma)\,.
\end{equation}
It was shown in \cite{Borunda} that Levi-Civita is a solution for the equation of the connection in any of these theories. As far as we know, the general solution remains unknown, but we have found that the action presents the projective symmetry, $\Gamma_{\mu\nu}{}^{\rho}\rightarrow\Gamma_{\mu\nu}{}^{\rho}+\mathcal{A}_{\mu}\delta_{\nu}^{\sigma}$. The proof is the following. Under the projective transformation, the Riemann tensor is modified:
\begin{equation}
R_{\mu\nu\rho}{}^{\lambda}\rightarrow R_{\mu\nu\rho}{}^{\lambda}+\mathcal{F}_{\mu\nu}\delta_{\rho}^{\lambda}\,.
\end{equation}
Then, the lagrangian transforms:
\begin{equation}
\mathcal{L}_{n}\rightarrow\delta_{\nu_{1}}^{[\mu_{1}}...\delta_{\nu_{2n}}^{\mu_{2n}]}\left(R_{\mu_{1}\mu_{2}}{}^{\nu_{1}\nu_{2}}+\mathcal{F}_{\mu_{1}\mu_{2}}g^{\nu_{1}\nu_{2}}\right)...\left(R_{\mu_{2n-1}\mu_{2n}}{}^{\nu_{2n-1}\nu_{2n}}+\mathcal{F}_{\mu_{2n-1}\mu_{2n}}g^{\nu_{2n-1}\nu_{2n}}\right)\,,
\end{equation}
and the $\nu$\textquoteright s antisymmetrization cancels all the terms proportional to the metric, so 
\begin{equation}
\delta_{\text{proj}}\mathcal{L}_{n}=0\,,\qquad\forall n\qquad\Rightarrow\qquad\delta_{\text{proj}}\left\{ S_{\text{Lov}}[g,\,\Gamma]+S_{\text{matter}}[g,\,\psi]\right\} =0\,.\,\,\text{(QED)}
\end{equation}
Consequently, since Levi-Civita ($\Gamma^{(g)}{}_{\mu\nu}{}^{\sigma}$) is a solution, then, $\Gamma^{(g)}{}_{\mu\nu}{}^{\sigma}+\mathcal{A}_{\mu}\delta_{\nu}^{\sigma}$ is also a solution. Indeed, the whole set of solutions can be separated into equivalence classes of projectively related connections. So if a new solution $\Gamma^{\text{sol}}{}_{\mu\nu}{}^{\sigma}$ of a Lovelock theory that has not the form $\Gamma^{(g)}{}_{\mu\nu}{}^{\sigma}+\mathcal{A}_{\mu}\delta_{\nu}^{\sigma}$ (for some $\mathcal{A}_{\mu}$) is found, then we could build a family of new solutions just adding a term $\mathcal{B}_{\mu}\delta_{\nu}^{\sigma}$ where $\mathcal{B}_{\mu}$ is arbitrary. This property also holds for any other lagrangians with projective invariance, such as $f(R)$ gravity.

Other theories we have tested are those with quadratic torsion corrections to the Einstein-Hilbert lagrangian. The torsion corrections we consider
are only those with even parity: 
\begin{align}
S[g,\,\Gamma,\,\psi] & =\frac{1}{2\kappa}\int g^{\mu\nu}R_{\mu\nu}(\Gamma)\,\sqrt{\left|g\right|}\mathrm{d}^{D}x\,\,+\,\,S_{\text{matter}}[g,\,\psi]\nonumber \\
 & \qquad+\frac{1}{2\kappa}\int\left[b_{1}T^{(1)}{}_{\mu\nu\rho}T^{(1)}{}^{\mu\nu\rho}+b_{2}T^{(2)}{}_{\mu\nu\rho}T^{(2)}{}^{\mu\nu\rho}+b_{3}T^{(3)}{}_{\mu\nu\rho}T^{(3)}{}^{\mu\nu\rho}\right]\sqrt{\left|g\right|}\mathrm{d}^{D}x\,,
\end{align}
where $b_{i}$ are arbitrary dimensionless real constants and $T^{(i)}{}_{\mu\nu\rho}$ are the irreducible components of the torsion (see \cite{McCrea}). For these extensions the equivalence between metric and Palatini formalism holds.

\section{Conclusions}

To conclude we summarize our results. We have seen that Einstein-Hilbert gravity in the Palatini formalism have some interesting features. If we couple this theory with a matter action through the metric (and not the connection) the result is physically indistinguishable from the dynamics obtained assuming Levi-Civita as the fundamental affine structure from the beginning (metric formalism). 

The general solution of the equation of the connection is Levi-Civita plus the term $\mathcal{A}_{\mu}\delta_{\nu}^{\sigma}$ where $\mathcal{A}_{\mu}$ is an undetermined field. However the equations of motion are the same as in metric formalism. Therefore, we get to different mathematical descriptions (related through the projective symmetry) that describe the same physics. In other words, it is not necessary to set the connection
to be Levi-Civita by hand. The dynamics fixes the affine structure.

Another additional property of the Palatini connections is that they are the only affine structures whose parallel transport is homothetic with respect to the Levi-Civita transport. So the directions obtained in both cases are coincident.

We have also proved that an autoparallel of a given Palatini connection is a trajectory with critical length (autoparallel of Levi-Civita). The undetermined field $\mathcal{A}_{\mu}$ for a free falling observer can be absorbed in a reparametrization of its worldline, so it has no physical meaning since a particular choice of the parameter is meaningless.

We have found solutions of this kind in more general theories, for example, in Lovelock gravity. In addition, if we admit matter that does not feel the connection, the equivalence between formalisms can be extended from Einstein-Hilbert to any theory with additional quadratic torsion terms in the action. Current work involves the treatment with more general matter and with additional terms that introduce, for example, dynamics for the torsion field.

\vspace{2mm}

\noindent
{\bf Acknowledgements}\\
We are grateful to A. Bernal and M. Sánchez for previous collaborations. We also wish to thank F.W. Hehl and J. Pons for useful comments and discussions. This work has been funded by the Junta de Andalucía (FQM101), the Universidad de Granada (PP2015-03), the MEC (FIS2016-78198-P) and the Unidad de Excelencia UCE-PP2016-02. In addition, J.A.O. was also supported by a PhD contract of the Plan Propio of the Universidad de Granada, and A.J.C. by a PhD contract of the FPU program (ref: FPU15/02864) of the MEC.


\end{document}